\documentclass[prd,nofootinbib,showpacs,superscriptaddress,preprint]{revtex4}
\usepackage[T1]{fontenc}
\usepackage{amsmath,amssymb}
\usepackage{epsfig}
\usepackage{dcolumn}
\usepackage{graphicx}
\usepackage[usenames,dvipsnames]{color}
\usepackage{slashed}
\usepackage[colorlinks,citecolor=blue]{hyperref}
\usepackage{pdfpages}
\usepackage{float}
\begin{document}
	\title{\boldmath Effect of sterile neutrino on low energy processes in minimal extended seesaw with $\Delta(96)$ symmetry and  $\text{TM}_{1}$ mixing}
	
	\author{Nayana Gautam}
	\email{nayana@tezu.ernet.in}
	\affiliation{Department of Physics, Tezpur University, Tezpur - 784028, India}
	
	\author{R. Krishnan}
	\email{krishnan@sinp.ernet.in}
	\affiliation{Saha Institute of Nuclear Physics,
		1/AF Bidhannagar, Kolkata 700064, India}
	
	\author{Mrinal Kumar Das}
	\email{mkdas@tezu.ernet.in}
	\affiliation{Department of Physics, Tezpur University, Tezpur - 784028, India}

\begin{abstract}
		
	We study the effect of sterile neutrino on some low scale processes in the framework of minimal extended seesaw (MES). MES is the extension of the seesaw mechanism with the addition of sterile neutrino of intermediate mass. The MES model in this work is based on $\Delta(96)\times C_{2}\times C_{3}$ flavor symmetry. The structures of mass matrices in the framework lead to $TM_{1}$ mixing with  $\mu \text{-}\tau$ symmetry. The model predicts maximal value of Dirac CP phase. We carry out our analysis to study the new physics contributions from the sterile neutrino to different charged lepton flavor violation (cLFV) processes involving muon and tau leptons as well as neutrinoless double beta decay (0$\nu\beta\beta$). The model predicts normal ordering (NO) of neutrino masses and we perform the numerical analysis considering normal ordering (NO) only. We find that sterile neutrino mass in GeV range can lead to cLFV processes that are within the reach of current and planned experiments. The GeV scale sterile neurtrino in our model is consistent with the current limits on the effective neutrino mass set by $0\nu\beta\beta$ experiments.  
	\end{abstract}
	\maketitle

\section{\label{sec:level1}Introduction}

The observed neutrino oscillation phenomenon, the origin of the idea behind the massive nature of neutrinos, has been one of the most appealing evidences to expect physics beyond standard model (BSM). Neutrino oscillation probabilities are dependent on the three mixing angles,the neutrino mass-squared differences ($\Delta m_{21}^{2}$, $\Delta m_{31}^{2}$) and the Dirac CP phase ($\delta_{CP}$). Though there are precise measurements of the mixing angles and mass squared differences, yet there are no conclusive remarks on ($\delta_{CP}$) or the mass ordering of the neutrinos. NO$\nu$A \cite{PhysRevD.98.032012} and T2K \cite{Abe:2015awa} experiments have recently provided hint towards the CP violation in Dirac neutrino matrix. Again, another important unsolved issue is the mass ordering of the neutrinos whether it is normal ($m_{1}<m_{2}<m_{3}$) or inverted ($m_{3}<m_{1}<m_{2}$). There are some other open questions in particle physics as well as cosmology such as CP violation in lepton sector, baryon asymmetry of the universe and particle nature of dark matter. Motivated by these shortcomings, different beyond standard model (BSM) theories \cite{Mohapatra:1980yp} are pursued in different experiments.

Many searches for new physics beyond standard model are going on in different experiments. Charged lepton flavor violating (cLFV) processes can provide a way to search for new physics beyond standard model. cLFV processes are heavily suppressed in the standard model . However, the well established neutrino oscillation phenomenon give a signal towards the flavor violation in the charged lepton sector also. There are present and planned experiments to search for lepton flavor violating radiative decay ($l_{i}\longrightarrow l_{j}\gamma$)\cite{Bernstein:2013hba} and also three body decays ($l_{i}\longrightarrow l_{j}l_{k}l_{k}$) \cite{Mihara:2013zna}. The present and future experimental constraints on cLFV processes can be found in table \ref{tab1a} and \ref{tab1b} . In this work, we study the transition among the three charged leptons. However, the transitions of muon such as $\mu-e,N$, $\mu\longrightarrow eee$, $\mu\longrightarrow e\gamma$ \cite{Lindner:2016bgg,Bertl:2006up} and recently proposed $\mu^{-}e^{-}\longrightarrow e^{-}e^{-}$ \cite{PhysRevLett.105.121601} are extensively analyzed as the parent particle is substantially available in the cosmic radiation as well as in dedicated accelerators\cite{Calibbi:2017uvl}. Many other challenging cLFV processes are those which involve the third family of leptons (taus) as it opens many flavor violating channels. Among these $\tau\longrightarrow e\gamma$, $\tau\longrightarrow \mu\gamma$, $\tau\longrightarrow 3e$ and $\tau\longrightarrow 3\mu$ are significant. The processes involving taus also open up many channels involving hadrons in the final state such as $\tau\longrightarrow l\pi^{0}$ , $\tau\longrightarrow l\pi^{+}\pi^{-}$\cite{Calibbi:2017uvl,Hayasaka:2010np} .

\begin{table}[H]
	\centering
	\begin{tabular}{|c|c|c|}
		
		\hline 
		cLFV Process & Present Bound & Future sensitivity \\ 
	    \hline
		$\mu\longrightarrow eee$	& $1.0\times10^{-12}$& $\sim 10^{-16}$  \\ 
		
		$\mu\longrightarrow e\gamma$ &$5.7\times10^{-13}$& $6.0\times10^{-14}$  \\ 
		
		$\tau\longrightarrow e\gamma$ &  $3.3\times10^{-8}$& $\sim 3\times10^{-9}$  \\ 
	
		$\tau\longrightarrow \mu\gamma$&  $4.4\times10^{-8}$& $\sim 10^{-9}$  \\ 
		
		$\tau\longrightarrow eee$& $2.7\times10^{-8}$& $\sim 10^{-9}$\\ 
		\hline  
	\end{tabular} 
	\caption{Current experimental bounds and future sensitivities for different cLFV
		processes.\cite{TheMEG:2016wtm,Aubert:2009ag,Bellgardt:1987du,Hayasaka:2010np}}\label{tab1a}
\end{table} 
\begin{table}[H]
	\centering
	\begin{tabular}{|c|c|}
		\hline 
		cLFV Process & Experimental Bound \\
	\hline
       $(\mu - e,Au)$ & $7\times10^{-13}$\\ 
       
      $(\mu - e,Al)$ & $3\times10^{-12}$\\ 
       \hline
	\end{tabular} 
\caption{Experimental bounds for the
	processes CR $(\mu - e,N)$. \cite{Bertl:2006up,Bellgardt:1987du}}\label{tab1b}
\end{table} 
There are various theoretical models which are the extension  of SM that can account for cLFV processes \cite{Deppisch:2004fa,Ilakovac:1994kj,Abada:2016vzu}. These models usually introduce new particle fields to act as a source of flavor violation. The models with heavy sterile neutrinos can provide prominent contributions to cLFV processes. There are many theoretical motivations as well as experimental background for the existence of sterile neutrinos. The anomalies of the LSND \cite{Athanassopoulos:1997pv} and MiniBooNE \cite{Aguilar-Arevalo:2018gpe} results provide a hint towards the presence of one or two sterile neutrino states. Again from the theoretical point of view, the addition of sterile fermions into the standard model can explain the neutrino mass and also mixing \cite{Naumov:2019kwm}. Moreover, sterile neutrino can account for many cosmological observations like dark matter \cite{Gautam:2019pce,Adhikari:2016bei,Dolgov:2000ew,Hamann:2011ge,Dodelson:1993je} and baryon asymmetry of universe (BAU) \cite{Lucente:2016vru,Gautam:2020wsd}. Furthermore, their mixing with the active neutrinos can contribute to certain non-oscillation processes like neutrino-less double decay (0$\nu\beta\beta$) amplitude or in beta decay spectra in the KATRIN experiment \cite{Abada:2018qok,Abazajian:2017tcc}.  
To study the effect of sterile neutrino on low scale processes, we have chosen minimal extended seesaw (MES) framework augmented with $\Delta(96)$ flavor symmetry. In MES framework, three right-handed neutrinos and one additional gauge singlet field S are added to the SM particle content \cite{Zhang:2011vh,Barry:2011wb}. The extra sterile state may have significant contribution to cLFV processes and 0$\nu\beta\beta$ depending on its mass and mixing with the active neutrinos in the model. In the present work, $C_{2}$ and $C_{3}$ discrete groups are introduced along with $\Delta(96)$ to avoid the unwanted couplings among the particles. The mass matrices constructed in the MES model embedded with $\Delta(96)$ flavor symmetry lead to a particular mixing pattern widely known as $\text{TM}_{1}$ mixing \cite{Luhn:2013lkn}. $\text{TM}_{1}$ mixing is one of the most significant mixing pattern which comply with the experimental predictions mixing angles and Dirac CP phase. In the present work, after constructing the mass matrices leading to $\text{TM}_{1}$ mixing, the model parameters have been evaluated using three neutrino oscillation parameters and then mass and mixing of the particles are calculated as a function of these model parameters. Further, we have evaluated different observables characterizing the different cLFV processes and neutrinoless double beta decay (0$\nu\beta\beta$). 

The paper is planned as follows. In section \ref{sec:level2}, we describe the $\text{TM}_{1}$ mixing and the model with $\Delta(96)$ flavor symmetry. The particles are assigned with different charges under the symmetry group and the mass matrices involved in the model are constructed. Section \ref{sec:level3} is the brief discussion of different cLFV processes and contribution of sterile neutrinos in such processes. In section \ref{sec:level4}, we briefly discuss about neutrinoless double beta decay process in presence of heavy sterile neutrinos. The  results of the numerical analysis are discussed in detail in section \ref{sec:level5}. Finally, we conclude in section \ref{sec:level6}.

\section{\label{sec:level2} Minimal Extended Seesaw with $\Delta(96)$ flavor symmetry with $\text{TM}_{1}$ mixing}
\subsection{The MES Framework}
Minimal Extended Seesaw (MES) is the extension of canonical type-I seesaw by the addition of extra gauge singlet field, $\nu_{s}$ to accommodate sterile neutrinos. This field has a coupling with the heavy right handed neutrino fields that are present in type-I seesaw \cite{Zhang:2011vh,Nath:2016mts,Das:2018qyt}. Thus the Lagrangian in this MES model can be obtained as \cite{Zhang:2011vh},
\begin{equation}\label{eq:1}
-\mathcal{L}  = \bar{\nu_{L}}M_{D} N + \frac{1}{2} N^{c}M_{R}N + \bar{S}M_{S}N + h.c
\end{equation} 
Subsequently, the mass matrix arising from the Lagrangian in Eq.(\ref{eq:1}) in the basis $(\nu_{L},N^{c},S^{c})$ can be written as,
\begin{equation}
M_{\nu}^{7\times7}=\left(\begin{array}{ccc}
0 & M_{D}& 0 \\
{M_{D}}^{T} & M_{R} &M_{S}^{T} \\
0 &  M_{S} & 0
\end{array}\right)
\label{massmatrix}
\end{equation}
Since the right-handed neutrinos are much heavier than the electroweak scale as in case of type-I seesaw, they should be decoupled at low scales. Effectively,
the full $7\times7$ matrix can be block diagonalised into a $4\times4$ neutrino
mass matrix as follows \cite{Zhang:2011vh},
\begin{equation}\label{massmatrix1}
M_{\nu}^{4\times4}=-\left(\begin{array}{ccc}
M_{D}{M_{R}}^{-1}{M_{D}}^{T} & M_{D}{M_{R}}^{-1}{M_{S}}^{T}  \\
M_{S}({M_{R}}^{-1})^{T}{M_{D}}^{T} & M_{S}{M_{R}}^{-1}{M_{S}}^{T} 
\end{array}\right)
\end{equation}
Assuming $M_{S}>M_{D}$, the active neutrino mass matrix of Eq.(\ref{massmatrix1}) takes the form as,
\begin{equation}\label{eq:1bb}
M_{\nu}\simeq M_{D}M_{R}^{-1}M_{S}^{T}(M_{S}M_{R}^{-1}M_{S}^{T})^{-1}M_{S}M_{R}^{-1}M_{D}^{T}-M_{D}M_{R}^{-1}M_{D}^{T}
\end{equation}
The sterile neutrino mass can be obtained as,
\begin{equation}\label{eq:1a}
m_{4}\simeq M_{S}M_{R}^{-1}M_{S}^{T}
\end{equation}
The charged lepton mass matrix in general can be diagonalised using unitary matrices $U_{L}$ and $U_{R}$ as follows ,
\begin{equation}\label{eq:1cc}
U_{L}M_{l}U_{R}^{\dagger}= \text{diag}(m_{e},m_{\mu},m_{\tau})
\end{equation}
Again, we obtain the light neutrino masses using unitary matrix $U_{\nu}$ as,
\begin{equation}\label{eq:1ee}
U_{\nu}^{\dagger}M_{\nu}^{3\times3}U_{\nu}= \text{diag}(m_{1},m_{2},m_{3})
\end{equation}
The $4\times4$ neutrino mixing matrix in MES model using  $U_{L}$ and $U_{\nu}$  can be obtained as \cite{Krishnan:2020xeq} ,
\begin{equation}
V =\left(\begin{array}{ccc}
U_{L}(1-\frac{1}{2} R R^{\dagger})U_{\nu} & U_{L} R \\
- R^{\dagger}U_{\nu} & 1-\frac{1}{2} R^{\dagger} R
\end{array}\right)
\label{massmatrix2}
\end{equation}

The matrix $U_{L}R$ governs the active-sterile mixing in which R can be expressed as,
\begin{equation}\label{eq:1b}
R = M_{D}M_{R}^{-1}M_{S}^{T}(M_{S}M_{R}^{-1}M_{S}^{T})^{-1}
\end{equation}
and,
\begin{equation}\label{eq:1c}
U_{L}R = \text{diag}(U_{e4},U_{\mu 4},U_{\tau 4})^{T}
\end{equation}
Finally, the $3\times3$ lepton mixing matrix (PMNS) can be written as \cite{Krishnan:2020xeq},
\begin{equation}\label{eq:1f}
U_{PMNS} = U_{L}(1-\frac{1}{2} R R^{\dagger})U_{\nu}
\end{equation}
\begin{equation}\label{eq:1ff}
U_{PMNS}\simeq U_{L}U_{\nu}
\end{equation}
Thus PMNS matrix can be obtained by multiplying the diagonalising matrix of charged lepton mixing matrix and that of the effective seesaw matrix. $U_{L}$ is identity matrix in the framework where charged lepton mass matrix is diagonal.
\subsection{$TM_{1}$ Mixing}
Trimaximal ($TM_{1}$) mixing is a mixing ansatz that preserves the first column of tri-bimaximal mixing $U_{TBM}$ and mixes its second and third columns. It is a perturbation to TBM mixing and we can write the mixing matrix as \cite{King:2019vhv,Luhn:2013lkn,Chakraborty:2020gqc},
\begin{equation}
U_{{TM_{1}}}=U_{\text{TBM}}\left(\begin{array}{ccc}
1&0& 0\\
0& \cos{\theta} & \sin{\theta}e^{-i\zeta} \\
0 & -\sin{\theta}e^{i\zeta}& \cos{\theta}
\end{array}\right)
\label{matrixTM}
\end{equation}
\begin{equation}
U_{{TM_{1}}}=\left(\begin{array}{ccc}
\frac{\sqrt{2}}{\sqrt{3}}&\frac{\cos{\theta}}{\sqrt{3}}& \frac{\sin{\theta}}{\sqrt{3}}e^{-i\zeta}\\
\frac{-1}{\sqrt{6}}&\frac{\cos{\theta}}{\sqrt{3}}-\frac{\sin{\theta}}{\sqrt{2}}e^{i\zeta}& \frac{\sin{\theta}}{\sqrt{3}}e^{-i\zeta}+\frac{\cos{\theta}}{\sqrt{2}}\\
\frac{-1}{\sqrt{6}}&\frac{\cos{\theta}}{\sqrt{3}}+\frac{\sin{\theta}}{\sqrt{2}}e^{i\zeta}& \frac{\sin{\theta}}{\sqrt{3}}e^{-i\zeta}-\frac{\cos{\theta}}{\sqrt{2}}
\end{array}\right)
\label{matrixTM1}
\end{equation}
Comparing above mixing matrix in Eq.(\ref{matrixTM1}) with the standard PMNS mixing matrix, one can obtain the three mixing angles in terms of $\theta$ as follows \cite{Krishnan:2019xmk}:
\begin{align}
\sin^2{\theta_{13}}&= \frac{\sin^2{\theta}}{3} \\
\sin^2{\theta_{23}}&= \frac{1}{2} \left(1+\frac{\sqrt{6}\sin{2\theta}\cos{\zeta}}{3-\sin^2{\theta}}\right)\\
\sin^2{\theta_{12}}&= 1- \frac{2}{3-\sin^2{\theta}}\\
J_{\text{CP}} & = \frac{\sin{2\theta}\sin{\zeta}}{6 \sqrt{6}} 
\end{align}
The Jarlskog’s rephasing invariant $J_{\text{CP}}$ can be written in terms of the elements of the mixing matrix as,
\begin{align}\label{eq:2b}
J_{\text{CP}}& = \text{Im}(U_{\mu 3}U_{e 3}^{\ast}U_{e 2}U_{\mu 2}^{\ast})\nonumber \\
&=\frac{1}{8}\text{sin} \delta \text{sin}2\theta_{12}\text{sin}2\theta_{23}\text{sin}2\theta_{13}\text{cos}\theta_{13} 
\end{align}
One can write the expression for the CP phase in the $TM_{1}$ scenario as,
\begin{equation}\label{eq:1d}
\sin^{2}\delta = \frac{8\text{sin}^{2}\theta_{13}(1-3\text{sin}^{2}\theta_{13})-\text{cos}^{4}\theta_{13}\text{cos}^{2}2\theta_{23}}{8\text{sin}^{2}\theta_{13}\text{sin}^{2}2\theta_{23}(1-3\text{sin}^{2}\theta_{13})}
\end{equation}
For a given $\theta_{13}(\theta)$, the $TM_{1}$ mixing with $\mu \text{-}\tau$ symmetry leads to maximal CP violation. Again it can be seen that if $\zeta = \pm \frac{\pi}{2}$ , $\theta_{23}= \frac{\pi}{4}$, which leads to$\mu \text{-}\tau$ symmetry.
\subsection{The Lagrangian}
In this work, we have used $\Delta (96)$ flavor symmetry \cite{Ding:2012xx,King:2012in,King:2013vna,Fonseca:2014koa} giving rise to unique textures of the mass matrices involved in the MES model. For a brief discussion about properties of $\Delta (96)$ ,its character table and tensor product rules please refer Appendix \ref{appen1}.  $\Delta (96)$ symmetry is further augmented by $C_{2}$ ans $C_{3}$ discrete flavor symmetries to get rid of some unwanted interactions. The particle assignments in the model are shown in table \ref{tab1}.

In our MES model, the lepton doublets of the SM and the SM gauge singlets transform as triplets $3_{i}$ and $\bar{3_{i}}$ of $\Delta (96)$ respectively. The sterile neutrino and the three right-handed charged leptons transform as singlets under this symmetry group. We introduce flavons $\phi_{\mu}$,$\phi_{\tau}$,$\phi_{S}$ transforming as triplets $3_{i}$ while $\phi_{M}$,$\phi_{D}$ are triplet $3^{\prime}$ and $\phi_{Mi}$,$\phi_{Di}$ are $\bar{3_{i}^{\prime}}$ under  $\Delta (96)$. These fields are also assigned various
charges under the $C_{2}$ and $C_{3}$ group which can be found in table \ref{tab1}
\begin{table}[H]
	\centering
	\begin{tabular}{|c|c|c|c|c|c|c|c|c|c|c|c|c|c|c|c|}
		
	\hline 
	Field	& $L$ & $e_{R}$ & $\mu_{R}$ & $\tau_{R}$ & N & S & $\phi_{\mu}$ &$\phi_{\tau}$ & $\phi_{M}$  & $\phi_{Mi}$ & $\phi_{D}$& $\phi_{Di}$ & $\phi_{S}$ \\ 
	\hline 
	$\Delta(96)$ &$3_{i}$  & $1$& $1$ & $1$ & $\bar{3_{i}}$& $1$ &$3_{i}$ &$3_{i}$ & $3^{\prime}$  &$\bar{3_{i}^{\prime}}$ &$3^{\prime}$   & $\bar{3_{i}^{\prime}}$ &$3_{i}$  \\
	\hline 
	$C_{3}$ & $1$ & $1$ &$\omega$& $\bar{\omega}$ & $1$ & $1$ &  $\bar{\omega}$ &$\omega$&$1$&$1$ & $1$ & $1$ &$1$ \\
	\hline 
	$C_{2}$& $1$ & $1$ & $1$ &$1$& $-1$ & $1$ & $1$ &  $1$ &$1$ &$1$& $-1 $ & $-1$ & $-1 $ \\
	\hline 
	$C_{3}$& $1$ & $1$ &$1$& $1$ & $1$ & $\omega$ &  $1$ &$1$ & $1$ & $1$& $1$ & $1$& $\bar{\omega}$  \\
	\hline 
	\end{tabular} 
	\caption{Fields and their respective transformations under the symmetry group of the model.} \label{tab1}
\end{table}
The Yukawa Lagrangian for the charged leptons and also for the neutrinos can be expressed as:
\begin{equation}
-\mathcal{L}  = \mathcal{L}_{\mathcal{M_{L}}}+\mathcal{L}_{\mathcal{M_{D}}} + \mathcal{L}_{\mathcal{M}}+ \mathcal{L}_{\mathcal{M_{S}}}+ h.c
\end{equation}
$\mathcal{L}_{\mathcal{M_{D}}} $ represents Dirac neutrino Lagrangian given as,
\begin{equation}\label{eq:3}
\mathcal{L}_{\mathcal{M_{D}}} =  \frac{y_{D}}{\Lambda}(\bar{L}N)_{3^{\prime}}\tilde{H} \phi_{D} + \frac{y_{Di}}{\Lambda}(\bar{L}N)_{3_{i}^{\prime}}\tilde{H} \phi_{Di}
\end{equation}
The neutrino Majorana mass term $\mathcal{L}_{\mathcal{M}}$ can be expressed as,  
\begin{equation}\label{eq:4}
\mathcal{L}_{\mathcal{M}} = y_{M}(\bar{N^{c}}N)_{3^{\prime}} \phi_{M} + y_{Mi}(\bar{N^{c}}N)_{3_{i}^{\prime}}\phi_{Mi}
\end{equation}
The interactions between the sterile and the right handed neutrinos are involved in $\mathcal{L}_{\mathcal{M_{S}}}$.
\begin{equation} \label{eq:5} 
\mathcal{L}_{\mathcal{M_{S}}} = y_{S}{\bar{{S}^{c}}}N\phi_{S}
\end{equation}
$ \mathcal{L}_{\mathcal{M_{L}}}$ is the Lagrangian for the charged leptons which can be written as
\begin{equation} \label{eq:6} 
\mathcal{L}_{\mathcal{M_{L}}} = \frac{y_{\mu}}{\Lambda}\bar{L}H\phi_{\mu}\mu_{R}+\frac{y_{\tau}}{\Lambda}\bar{L}H\phi_{\tau}\tau_{R} +\frac{y_{e}}{\Lambda^{2}}\bar{L}H(\bar{\phi_{\tau}}\bar{\phi_{\mu}})_{3_{i}} e_{R}
\end{equation}
After Spontaneous Symmetry Breaking (SSB) , the scalar fields acquire VEV's which are assigned as:
\begin{equation} \label{eq:6a} 
\langle \phi_{\mu} \rangle = v_{\mu}(1,\bar{\omega},\omega)^{T}, \; \langle \phi_{\tau} \rangle = v_{\tau}(1,\omega,\bar{\omega})^{T}, \;\langle \phi_{S} \rangle = (0,v_{S},-v_{S})$$ 
$$\langle\phi_{M} \rangle = v_{M}(1,1,1)^{T},\; \langle\phi_{Mi} \rangle = v_{Mi}(1,0,-1)^{T},\; \langle\phi_{D} \rangle = v_{D}(0,1,0)^{T},\; \langle\phi_{Di} \rangle = v_{Di}(1,0,-1)^{T}
\end{equation}

\subsection{The Mass Matrices involved in the Model}
The textures of the mass matrices involved in MES model can be obtained using flavon alignments defined with residual symmetries under our flavor group. With these flavon alignments mentioned above, we obtain the charged-lepton and the neutrino mass matrices.
In the charged lepton sector, $\bar{L}$ which couples to $l_{R} (l= e,\mu,\tau)$ through the flavon $\phi_{\mu}$ and $\phi_{\tau}$. Using the VEV's of the flavons and the Higgs in the Lagrangian given by Eq.(\ref{eq:6}) ,the charged lepton mass matrix can be written as,
\begin{equation} \label{eq:u1} 
M_{C}= \frac{i \sqrt{3}v v_{\mu}v_{\tau}}{\Lambda^{2}}\left(\begin{array}{ccc}
y_{e} & 0 &  0\\
y_{e} & 0 & 0\\ 
y_{e} & 0 & 0
\end{array}\right) + \frac{v}{\Lambda}\left(\begin{array}{ccc}
0 & y_{\mu}v_{\mu} &y_{\tau}v_{\tau} \\
0 & \bar{\omega} y_{\mu}v_{\mu} & \omega y_{\tau}v_{\tau}\\ 
0 & \omega y_{\mu}v_{\mu} & \bar{\omega} y_{\tau}v_{\tau}
\end{array}\right)
\end{equation}
The charged lepton mass matrix $M_{C}$ is diagonalised using the unitary matrix $U_{L}$ given as,
\begin{equation} \label{eq:u6} 
U_{L} = \frac{1}{\sqrt{3}}\left(\begin{array}{ccc}
1 & 1 & 1 \\
1 & \omega & \bar{\omega}\\ 
1 &\bar{\omega} & \omega
\end{array}\right)
\end{equation}
$U_{L}$ is referred to as the $3\times3$ trimaximal matrix (TM) or the magic matrix.
\begin{equation} \label{eq:u7} 
U_{L} M_{C}\text{diag}(-i,1,1)  = \text{diag}(m_{e},m_{\mu},m_{\tau})
\end{equation}
and we obtain the masses of the charged leptons as,
\begin{equation} \label{eq:u8} 
m_{e} = 3 y_{e}v \frac{v_{\mu}v_{\tau}}{\Lambda^{2}} ,  m_{\mu} = \sqrt{3} y_{\mu}v \frac{v_{\mu}}{\Lambda} , m_{\tau} = \sqrt{3} y_{\tau}v \frac{v_{\tau}}{\Lambda}
\end{equation}
It is seen from Eq.(\ref{eq:u8}) that the mass scale of electron is suppressed by an additional factor $\frac{1}{\Lambda}$ compared to tau or muon mass similar to Froggatt-Nielsen mechanism of obtaining the mass hierarchy.

Again, from Eq.(\ref{eq:3}), we obtain the Dirac neutrino mass matrix as,
\begin{equation} \label{eq:u2} 
M_{D}= \frac{v}{\Lambda}\left(\begin{array}{ccc}
0 & -y_{Di}v_{Di} & 0 \\
-y_{Di}v_{Di} & y_{D}v_{D} & y_{Di}v_{Di}\\ 
0 & y_{Di}v_{Di} & 0
\end{array}\right)
\end{equation}
Denoting $\frac{y_{D}v_{D}v}{\Lambda}= m_{D}$ and $\frac{y_{Di}v_{Di}}{y_{D}v_{D}}=r_{1}$, we rewrite the Dirac mass matrix in Eq.(\ref{eq:u2}) as,
\begin{equation} \label{eq:u22} 
M_{D}= m_{D} \left(\begin{array}{ccc}
0 & -r_{1} & 0 \\
-r_{1} & 1 & r_{1}\\ 
0 & r_{1} & 0
\end{array}\right)
\end{equation}
$m_{D}$ has the dimension of mass similar to the order of the SM fermion masses and $ r_{1}$ is dimensionless.
The Majorana mass matrix for the heavy right-handed neutrinos can be obtained using the VEV's of $\phi_{M}$ and $\phi_{Mi}$ in Eq.(\ref{eq:4}) as,
\begin{equation} \label{eq:u3} 
M_{R}= \left(\begin{array}{ccc}
y_{M}v_{M} & -y_{Mi}v_{Mi} & 0 \\
-y_{Mi}v_{Mi} &y_{M}v_{M} & y_{Mi}v_{Mi}\\ 
0 & y_{Mi}v_{Mi} & y_{M}v_{M}
\end{array}\right)
\end{equation}
Here also, we denote $y_{M}v_{M}= m_{R}$ and $\frac{y_{Mi}v_{Mi}}{y_{M}v_{M}}= r_{2}$ and rewrite the above matrix as, 
\begin{equation} \label{eq:u33} 
M_{R}= m_{R}\left(\begin{array}{ccc}
1 & -r_{2} & 0 \\
-r_{2} &1 & r_{2}\\ 
0 & r_{2}& 1
\end{array}\right)
\end{equation}
$m_{R}$ has the dimension of mass at the scale of flavon VEV and $ r_{2}$ is dimensionless.

Finally, we obtain the mass matrix representing the coupling between right handed neutrinos and sterile neutrino as,
\begin{equation} \label{eq:u4} 
M_{S}= y_{S}v_{S} \left(\begin{array}{ccc}
0 & 1 & -1
\end{array}\right)
\end{equation}
or we can rewrite it as,
\begin{equation} \label{eq:u44}
M_{S}= m_{S} \left(\begin{array}{ccc}
0 & 1 & -1
\end{array}\right)
\end{equation}
where, $m_{S} = y_{S}v_{S}$ has the dimension of mass.

The light neutrino mass matrix in the framework of MES arising from the mass matrices in Eqs.(\ref{eq:u22},\ref{eq:u33},\ref{eq:u44}) can be written using Eq.(\ref{massmatrix1}) as:
\begin{equation} \label{eq:15} 
M_{\nu}=\left(\begin{array}{ccc}
K_{1} & -K_{2} & -K_{1} \\
-K_{2} &K_{3}&K_{2}\\ 
-K_{1} &K_{2}& K_{1}
\end{array}\right)
\end{equation}
where,
\begin{equation}\label{eq:15a}
K_{1} = -\frac{m_{D}^{2}r_{1}^{2}}{m_{R}(2+ r_{2}-r_{2}^{2})}
\end{equation}
\begin{equation}\label{eq:15b}
K_{2} = \frac{m_{D}^{2}r_{1}(-1+r_{1}(-1+r_{2}))}{m_{R}(2+ r_{2}-r_{2}^{2})}
\end{equation}

\begin{equation}\label{eq:15c}
K_{3} = -\frac{m_{D}^{2}(1+3 r_{1}^{2} - 2 r_{1}(-1+r_{2}) )}{m_{R}(2+ r_{2}-r_{2}^{2})}
\end{equation}

The effective seesaw mass matrix in Eq.(\ref{eq:15}) can be diagonalised in two steps using the unitary matrix $U_{BM}$ and $U_{\theta}$ as,
\begin{equation}\label{eq:16} 
U_{\theta}^{T}U_{BM}^{T}M_{\nu}U_{BM}U_{\theta}= \text{diag}(m_{1},m_{2},m_{3})
\end{equation}
or one may write,
\begin{equation}\label{eq:17a} 
M_{\nu}= U_{BM}U_{\theta}\text{diag}(m_{1},m_{2},m_{3})U_{\theta}^{T}U_{BM}^{T}
\end{equation}
The matrix $U_{\theta}$ and the bimaximal matrix $U_{BM}$ in the Eq.(\ref{eq:16}) are given by,
\begin{equation} \label{eq:17} 
U_{\theta}= \left(\begin{array}{ccc}
1 & 0 &0\\
0 & \text{cos}\theta & \text{sin}\theta\\ 
0 &-\text{sin}\theta&\text{cos}\theta
\end{array}\right),\;  U_{BM} = \left(\begin{array}{ccc}
\frac{1}{\sqrt{2}} & 0 & -\frac{1}{\sqrt{2}} \\
0 & 1 & 0\\ 
\frac{1}{\sqrt{2}} & 0 &\frac{1}{\sqrt{2}}
\end{array}\right)
\end{equation}
Comparing Eq.(\ref{eq:16}) with Eq.(\ref{eq:1ee}), we can write the neutrino mixing matrix $U_{\nu}$ as,
\begin{equation}\label{eq:18} 
U_{\nu} = U_{BM}U_{\theta}
\end{equation}
Therefore,using Eq.(\ref{eq:1ff}),the PMNS matrix in this model can be expressed as,
\begin{equation}\label{eq:19} 
U_{PMNS} \simeq U_{L}U_{BM}U_{\theta}
\end{equation}
Here, $U_{L}U_{BM}$ is the tri-bimaximal (TBM) mixing matrix,$U_{TBM}$. The multiplication of $U_{TBM}$ and $U_{\theta}$ mixes the 2nd and the 3rd columns of $U_{TBM}$ resulting in $TM_{1}$ mixing matrix $U_{TM1}$. Our construction of $M_{\nu}$ given in Eq.(\ref{eq:15}) leading to $TM_{1}$ mixing implies that $m_{1}= 0$, which rules out inverted hierarchy. Using this in Eq.(\ref{eq:17a}) and comparing with Eq.(\ref{eq:15}), we can find the expressions for model parameters $K_{1}$,$K_{2}$ and $K_{3}$ in terms of the parameters $\theta$, $m_{2}$ and $m_{3}$ as,
\begin{equation}
K_{1} = \frac{1}{2}(m_{3}\text{cos}^{2}\theta+ m_{2}\text{sin}^{2}\theta)
\end{equation}
\begin{equation}
K_{2} = \frac{1}{\sqrt{2}}(m_{3}-m_{2})\text{cos}\theta\text{sin}\theta
\end{equation}
\begin{equation}
K_{3} = m_{2}\text{cos}^{2}\theta+ m_{3}\text{sin}^{2}\theta
\end{equation}
\subsection{Sterile Neutrino Mass and Mixing in the Model}
Apart from the active neutrinos, the mass and mixing of the sterile neutrino present in the model play crucial role in cLFV processes which will be discussed in the next section. As mentioned above, the sterile neutrino mass can be obtained using Eq.(\ref{eq:1a}) and we can write the mass term for sterile neutrino as,
\begin{equation}\label{eq:d1} 
m_{4} = \frac{m_{S}^{2}(-2-2 r_{2} +2 r_{2}^{2})}{m_{R}(-1+2 r_{2}^{2})}
\end{equation}
The active-sterile mixing using Eq.(\ref{eq:1b}) and Eq.(\ref{eq:1c}) can be obtained as,
\begin{equation}\label{eq:d2}
U_{e4} =  \frac{m_{D}(-1+r_{1}-r_{2} +2 r_{1} r_{2})}{\sqrt{3}m_{S}(-2-2 r_{2}+2 r_{2}^{2})}
\end{equation}
\begin{equation}\label{eq:d3}
U_{\mu4} = \frac{m_{D}((1- i\sqrt{3})(1+ r_{2})+r_{1}(2+ 2 i \sqrt{3}+r_{2} +3i \sqrt{3}r_{2}))}{2 \sqrt{3} m_{S}(-2-2 r_{2}+2 r_{2}^{2})}
\end{equation}
\begin{equation}\label{eq:d4}
U_{\tau4} = \frac{m_{D}((1+ i\sqrt{3})(1+ r_{2})+r_{1}(2-2 i \sqrt{3}+r_{2} -3i \sqrt{3}r_{1}) )}{2 \sqrt{3} m_{S}(-2-2 r_{2}+2 r_{2}^{2})}
\end{equation}
In the above Eqs.(\ref{eq:d1},\ref{eq:d2},\ref{eq:d3},\ref{eq:d4}), $m_{D}$,$m_{R}$,$r_{1}$ and $r_{2}$ are the model parameters.
\section{\label{sec:level3}Charged Lepton Flavor Violating Processes}
\subsection{\label{sec:level3.1}Processes involving Muonic atoms}
Many on-going experiments like MECO, SINDRUM II \cite{Bertl:2006up},COMET \cite{Cui:2009zz} are involved in searching for $\mu-e$ conversion with different targets. The observable characterizing this process is  defined as ,
\begin{equation}
CR (\mu-e,N) = \frac{\Gamma(\mu^{-}+ N\rightarrow e^{-}+N)}{\Gamma(\mu^{-}+ N\rightarrow \text{all capture})}
\end{equation}
These experiments are running with different targets like Titanium (Ti), Lead (Pb), Gold (Au) Aluminum (Al) and give bounds for different targets. There are also some planned future experiments like the second phase of COMET experiment, Mu2e \cite{Carey:2008zz} to improve the sensitivity to this cLFV process. \\

There are several theoretical models to account for such rare LFV processes. As explained in \cite{Abada:2015oba}, in the extension of standard model with one heavy sterile neutrino, such processes originate from one-loop diagrams involving active and sterile neutrinos with non zero mixing angles. In the MES model,the conversion ratio can be written as \cite{Abada:2015oba},
\begin{equation}
CR (\mu-e,N) = \frac{2 G_{F}^{2}\alpha_{\omega}^{2}m_{\mu}^{5}} {(4\pi)^{2}\Gamma_{cap}(Z)}\mathrel{\Big|}4V^{(p)}(2\tilde{F_{u}^{\mu e}}+\tilde{F_{d}^{\mu e}})+4 V^{(n)}(\tilde{F_{u}^{\mu e}}+2\tilde{F_{d}^{\mu e}})+ DG_{\gamma}^{\mu e}\frac{s_{\omega}^{2}}{2\sqrt{4\pi\alpha}}\mathrel{\Big|}^{2}
\end{equation} 
In the above expression, $G_{F}$,$s_{\omega}$,$\Gamma_{cap}(Z)$ are Fermi constant, sine of weak mixing angle and capture rate of the nucleus respectively. Here, $\alpha=\frac{e^{2}}{4\pi}$ and $\tilde{F}_{q}^{\mu e}$ are form factors given as,
\begin{equation}
\tilde{F}_{q}^{\mu e}= Q_{q}s_{\omega}^{2} F_{\gamma}^{\mu e}+F_{Z}^{\mu e}(\frac{I_{q}^{3}}{2}- Q_{q}s_{\omega}^{2})+\frac{1}{4}F_{Box}^{\mu e qq}
\end{equation}
Here, $Q_{q}$ represents the quark electric charge which is $\frac{2}{3}$ and  $-\frac{1}{3}$ for up and down quark respectively. The weak isospin $I_{q}^{3}$ is $\frac{1}{2}$ and $-\frac{1}{2}$ for up and down quark respectively. The numerical values of $V^{(p)}$,$V^{(n)}$ and D in \cite{Ilakovac:1994kj}. In the small limit of masses ($x_{j}= \frac{m_{\nu j}^{2}}{m_{W}^{2}}\ll 1$), the form factors can be written as \cite{Abada:2015oba},
\begin{equation}\label{eq:11} 
F_{\gamma}^{\mu e} \rightarrow  \sum_{j=1}^{3+ n_{S}} U_{ej} U_{\mu j}^{ \ast}[-x_{j}]
\end{equation}
\begin{equation}\label{eq:12} 
G_{\gamma}^{\mu e} \rightarrow \sum_{j=1}^{3+ n_{S}} U_{ej}U_{\mu j}^{ \ast}[\frac{x_{j}}{4}]
\end{equation}
\begin{equation}\label{eq:13} 
F_{Z}^{\mu e} \rightarrow \sum_{j=1}^{3+ n_{S}} U_{ej}U_{\mu j}^{ \ast}[x_{j}(-\frac{5}{2}-ln x_{j})]
\end{equation}
\begin{equation}\label{eq:14} 
F_{Box}^{\mu eee} \rightarrow \sum_{j=1}^{3+ n_{S}} U_{ej} U_{\mu j}^{ \ast}[2 x_{j}(1+ln x_{j})]
\end{equation}   

There may be flavour violating non-radiative decay of $\mu^{-}$ into three electrons ($\mu\longrightarrow eee$) \cite{Kitano:2002mt}. Mu3e experiment running at PSI aims at finding the signatures of this type of decay \cite{Willmann:1998gd}. The branching ratio of this decay process can be written as,
\begin{align}
BR(\mu\longrightarrow eee)& = \frac{\alpha_{\omega}^{4}} {24576 \pi^{3}}\frac{m_{\mu}^{4}} {m_{W}^{4}}\frac{m_{\mu}} {\Gamma_{\mu}}2\mathrel{\Big|}\frac{1}{2} F_{Box}^{\mu eee}+F_{Z}^{\mu e}-2 s_{\omega}^{2}(F_{Z}^{\mu e}-F_{\gamma}^{\mu e})\mathrel{\Big|}^{2}+ 4 s_{\omega}^{4}|F_{Z}^{\mu e}- F_{\gamma}^{\mu e}|^{2} \nonumber \\
& + 16 s_{\omega}^{2} Re[(F_{Z}^{\mu e}+ \frac{1}{2} F_{Box}^{\mu eee})G_{\gamma}^{\mu e\ast}]-48 s_{\omega}^{4} Re[(F_{Z}^{\mu e}- F_{\gamma}^{\mu e})G_{\gamma}^{\mu e\ast}]\nonumber \\
&+ 32 s_{\omega}^{4}|G_{\gamma}^{\mu e}|^{2}[ln\frac{m_{\mu}^{2}}{m_{e}^{2}}-\frac{11}{4}]
\end{align}
In the above equation, the form factors can be obtained from Eq.(\ref{eq:11}) to Eq.(\ref{eq:14}).

The MEG experiment \cite{Adam:2013mnn} aims at investigating LFV process $\mu\longrightarrow e \gamma$ and there are many planned projects in search for this kind of decay. In the framework of minimal extended seesaw, the heavy neutrinos can cause $\mu\longrightarrow e \gamma$ decay. The branching ratio of the process can be given as,
\begin{equation}
BR(\mu\longrightarrow e\gamma) = \frac{\alpha_{\omega}^{3}s_{\omega}^{2}} {256 \pi^{2}}\frac{m_{\mu}^{4}} {M_{W}^{4}}\frac{m_{\mu}} {\Gamma_{\mu}}|G_{\gamma}^{\mu e}|^{2}
\end{equation}  
In the above equation,the total decay width of muon ($\Gamma_{\mu}$) is obtained as,
\begin{equation}
\Gamma_{\mu} = \frac{G_{F}^{2}m_{\mu}^{5}}{192 \pi^{3}}(1-8 \frac{m_{e}^{2}}{m_{\mu}^{2}})[1+\frac{\alpha_{em}}{2\pi}(\frac{25}{4}-\pi^{2})]
\end{equation}

Another possible cLFV process is the decay of a bound $\mu^{-}$ in a muonic atom into a pair of electrons $(\mu^{-}e^{-}\longrightarrow e^{-}e^{-})$ proposed by \cite{PhysRevLett.105.121601}. This particular decay process offers several advantages over three body decay processes from the experimental point of view. There are different classes of extension of SM which can show a contribution to such processes. In this model with one extra sterile state, the effective Lagrangian describing this process contains long range interactions and local interaction terms. The branching ratio of such process in muonic atoms,with an atomic number Z can be expressed as, 
\begin{align}
BR(\mu^{-}e^{-}\longrightarrow e^{-}e^{-},N)& = 24 \pi f_{Coul}(Z)\alpha_{\omega}\frac{m_{e}^{3}} {m_{\mu}^{3}}\frac{\tilde{\tau}_{\mu}} {\tau_{\mu}}(16|\frac{1}{2} (\frac{g_{\omega}}{4 \pi})^{2} (\frac{1}{2}F_{Box}^{\mu eee}+F_{Z}^{\mu e}-2 s_{\omega}^{2}(F_{Z}^{\mu e}-F_{\gamma}^{\mu e}))|^{2}\nonumber \\
&+4|\frac{1}{2} (\frac{g_{\omega}}{4 \pi})^{2}2 s_{\omega}^{2} (F_{Z}^{\mu e}- F_{\gamma}^{\mu e})|^{2} 
\end{align}
Here, $\tau_{\mu}$ represents the lifetime of free muon and the lifetime $\tilde{\tau}_{\mu}$ depends on specific elements. In our analysis, we have considered Al and Au in which value of $\tilde{\tau}_{\mu}$ are $8.64\times10^{-7}$ and $7.26\times10^{-8}$ respectively. This decay process would possibly be probed in the COMET collaboration. As suggested in many literature, we have used the future sensitivity of $CR (\mu-e,N)$ to constrain such decay process.

\subsection{\label{sec:level3.2}Processes involving Tau leptons}
There are many flavor violating channels open for tau lepton decays. Search for such decays involving taus is also challenging. Theoretical models which predict cLFV in the muon indicate a violation in the tau sector also. However, the amplitude of the process involving tau channel is enhanced by several order of magnitude in comparison to muon decays. Experiments like BaBar \cite{PhysRevLett.104.021802} and Belle \cite{Miyazaki:2011xe} provide limits to cLFV decays involving tau leptons. In this work, we have investigated three processes involving tau leptons $\tau\longrightarrow e\gamma$, $\tau\longrightarrow \mu\gamma$ and $\tau\longrightarrow eee$. The branching ratios of these mentioned process can be written as \cite{Ilakovac:1994kj},
\begin{equation}\label{eq:10a}
BR(\tau\longrightarrow e\gamma) = \frac{\alpha_{\omega}^{3}s_{\omega}^{2}} {256 \pi^{2}}\frac{m_{\tau}^{4}} {m_{W}^{4}}\frac{m_{\tau}} {\Gamma_{\tau}}|G_{\gamma}^{\tau e}|^{2}
\end{equation}  
\begin{equation}\label{eq:10b}
BR(\tau\longrightarrow \mu\gamma) = \frac{\alpha_{\omega}^{3}s_{\omega}^{2}} {256 \pi^{2}}\frac{m_{\tau}^{4}} {m_{W}^{4}}\frac{m_{\tau}} {\Gamma_{\tau}}|G_{\gamma}^{\tau\mu}|^{2}
\end{equation}
In the above equations, $\Gamma_{\tau}$ represents the total width of tau leptons with experimental value $\Gamma_{\tau}= 2.1581\times 10^{-12}$ GeV \cite{Ilakovac:1994kj}.
\begin{align}\label{eq:10c}
BR(\tau\longrightarrow eee)& = \frac{\alpha_{\omega}^{4}} {24576 \pi^{3}}\frac{m_{\tau}^{4}} {m_{W}^{4}}\frac{m_{\tau}} {\Gamma_{\tau}}2\mathrel{\Big|}\frac{1}{2} F_{Box}^{\tau eee}+F_{Z}^{\tau e}-2 s_{\omega}^{2}(F_{Z}^{\tau e}-F_{\gamma}^{\tau e})\mathrel{\Big|}^{2}+ 4 s_{\omega}^{4}|F_{Z}^{\tau e}- F_{\gamma}^{\tau e}|^{2} \nonumber \\
& + 16 s_{\omega}^{2} Re[(F_{Z}^{\tau e}+ \frac{1}{2} F_{Box}^{\tau eee})G_{\gamma}^{\tau e\ast}]-48 s_{\omega}^{4} Re[(F_{Z}^{\tau e}- F_{\gamma}^{\tau e})G_{\gamma}^{\tau e\ast}]\nonumber \\
&+ 32 s_{\omega}^{4}|G_{\gamma}^{\tau e}|^{2}[ln\frac{m_{\tau}^{2}}{m_{e}^{2}}-\frac{11}{4}]
\end{align} 
where, the composite form factors $F_{\gamma}^{\tau e}$,$G_{\gamma}^{\tau e}$,$F_{Z}^{\tau e}$ and $F_{Box}^{\tau eee}$ can be defined as follows:
\begin{equation}
F_{\gamma}^{\tau e} \rightarrow \sum U_{ej} U_{\tau j}^{ \ast}[-x_{j}]
\end{equation}
\begin{equation}
G_{\gamma}^{\tau e} \rightarrow \sum U_{ej}U_{\tau j}^{ \ast}[\frac{x_{j}}{4}]
\end{equation}
\begin{equation} 
F_{Z}^{\tau e} \rightarrow \sum U_{ej}U_{\tau j}^{ \ast}[x_{j}(-\frac{5}{2}-ln x_{j})]
\end{equation}
\begin{equation}
F_{Box}^{\tau eee} \rightarrow \sum U_{ej} U_{\tau j}^{ \ast}[2 x_{j}(1+ln x_{j})]
\end{equation} 
\section{\label{sec:level4} Neutrinoless Double Beta Decay (0$\nu\beta\beta$)}
The presence of sterile neutrinos in addition to the standard model particles may lead to new contributions to lepton number violating interactions like neutrinoless double beta decay(0$\nu\beta\beta$)\cite{Benes:2005hn,Awasthi:2013we,Borgohain:2018lro}. We have studied the contributions of the sterile state to the effective electron neutrino majorana mass $m_{\beta\beta}$ \cite{Abada:2018qok,Blennow:2010th}. The most stringent bounds on the effective mass by provided by KamLAND-ZEN experiment \cite{KamLAND-Zen:2016pfg}.
\begin{equation}\label{eq:20}
m_{\beta\beta} < 0.061- 0.165 eV
\end{equation}
The amplitude of these processes depends upon the neutrino mixing matrix elements and the neutrino masses. The decay width of the process is proportional to the effective electron neutrino majorana mass $m_{\beta\beta}$ which is in the case of standard contribution i.e. in the absence of any sterile neutrino is given as ,
\begin{equation}
m_{\beta\beta} = \mathrel{\Big|}\sum_{i = 1}^{3}{U_{ei}}^{2}m_{i}\mathrel{\Big|}
\end{equation}
The above equation is modified with the addition of sterile fermions and is given by \cite{Abada:2018qok} 
\begin{equation}\label{eq:21}
m_{\beta\beta} =\mathrel{\Big|}\sum_{i = 1}^{3}{U_{ei}}^{2}m_{i} + {U_{e4}}^{2}m_{4}\mathrel{\Big|}
\end{equation}
where, $m_{4}$ and ${U_{e4}}$ represent the mass and mixing of the sterile neutrino to the electron neutrino respectively.

\section{\label{sec:level5} Results of Numerical Analysis and Discussions}
It is evident from the above discussion that the neutrino mass matrix in Eq. (\ref{eq:15}) contains three model parameters $K_{1}$,$K_{2}$,$K_{3}$. We can express the experimentally measured six oscillation parameters $\textstyle {\Delta m_{21}^{2}}$, $\textstyle \Delta m_{31}^{2}$, $\textstyle \sin^{2}\theta_{12}$, $\textstyle \sin^{2}\theta_{23}$, $\sin^{2}\theta_{13}$, $\textstyle\delta_{CP}$ in terms of these model parameters. Hence, the three model parameters can be evaluated by comparing with the three oscillation parameters in $3\sigma$ range as given in table \ref{tab3} and then constrain the other parameters. These parameters  $\textstyle K_{1}$, $\textstyle K_{2}$, $\textstyle K_{3}$ in turn are related to $m_{D}$,$m_{R}$,$r_{1}$ and $r_{2}$ as given in Eqs.(\ref{eq:15a},\ref{eq:15b},\ref{eq:15c}) which are functions of Yukawa couplings and VEV's of the scalars. In our model, we have evaluated the model parameters comparing with experimental range of $\textstyle \Delta m_{21}^{2}$, $\textstyle \Delta m_{31}^{2}$, $\textstyle \sin^{2}\theta_{13}$. Since the lightest neutrino mass is zero in MES model, hence $\textstyle \Delta m_{21}^{2}$ and $\textstyle \Delta m_{31}^{2}$ will correspond the other two masses. Our construction of MES model with $TM_{1}$ mixing rules out the inverted ordering (IO) of the neutrino masses. The inverted ordering is disfavored with a $\textstyle\Delta\chi^{2} = 4.7$ \cite{Esteban:2018azc}. Hence, our results is in good agreement with the latest global data. Fig \ref{fig1a} represents correlation of different neutrino oscillation parameters with the model parameters.
\begin{table}[H]
	\centering
	\begin{tabular}{|c|c|}
		
		\hline 
		Oscillation parameters	& 3$\sigma$(NO) \\ 
		\hline 
		$\frac{\Delta m_{21}^{2}}{10^{-5}eV^{2}}$	& 6.80 - 8.02   \\ 
		
		$\frac{\Delta m_{31}^{2}}{10^{-3}eV^{2}}$	& 2.40 - 2.60  \\ 
		
        $\sin^{2}\theta_{13}$ &  0.0198 - 0.0243 \\ 
		\hline 
	\end{tabular} 
	\caption{Latest Global fit neutrino oscillation Data.}\label{tab3}
\end{table} 
\begin{figure}[H]
	\begin{center}
		\includegraphics[width=0.45\textwidth]{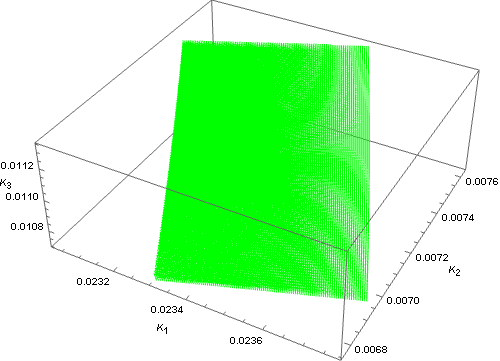}
	\end{center}
	\begin{center}
		\caption{Correlation plots for the model parameters (in eV).}
		\label{figmodel}
	\end{center}
\end{figure}
$TM_{1}$ with $\mu \text{-}\tau$ symmetry fixes the atmospheric mixing angle $\theta_{23}$ be $\frac{\pi}{4}$ i.e maximal atmospheric mixing angle. We have seen the predictions of the model on  Jarlskog parameter $J_{CP}$ and the Dirac CP phase $\delta_{CP}$ by evaluating these two parameters using Eq.(\ref{eq:2b}). The model predicts maximal $\delta_{CP}$ which is consistent with the current global fit. We have also calculated the sum of the three light neutrino masses from the model parameters. It predicts $\sum m_{i}$ within the range ($0.057-0.059$) which is below the cosmological upper bounds. Thus it is clear that the predictions of the model comply with the latest neutrino and cosmology data.

\begin{figure}[H]
	\begin{center}
		\includegraphics[width=0.85\textwidth]{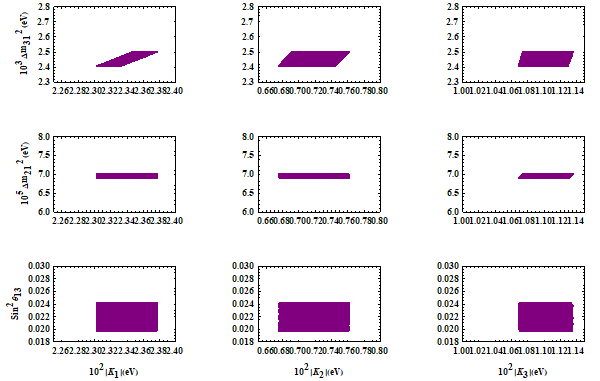}
	
	\end{center}
	\begin{center}
		\caption{The allowed region of $\Delta m^{2}_{31}$,$\Delta m^{2}_{21}$ and mixing angle $\sin^{2}\theta_{13}$ as a function of model parameters.}
		\label{fig1a}
	\end{center}
\end{figure}

\begin{table}[H]
	\centering
	\begin{tabular}{|c|c|c|}
		
		\hline 
		Parameters	& Predictions (NH) & Experimental Range \\ 
		\hline 
	
		$m_{\beta\beta}$ &  (0.014 - 0.016)eV  & <0.06 eV \\ 
		
         $\sum m_{i}$ & (0.057 - 0.059)eV &  <0.11 eV\\ 
		\hline 
	\end{tabular} 
	\caption{Predictions of the model on different parameters. The value of $m_{\beta\beta}$ is taken from KamLAND-ZEN experiment \cite{KamLAND-Zen:2016pfg} and $\sum m_{i}$ from latest Planck data \cite{Aghanim:2018eyx}.}\label{tab4}
\end{table} 
Apart from studying active neutrino phenomenology, we have calculated different observables related to the different cLFV processes with the numerically evaluated model parameters. All the masses and mixing in the model are dependent on the model parameters which are highly constrained from the neutrino oscillation data. The masses and mixing of the active as well as sterile neutrinos in turn are related to the observables of different cLFV processes and also $0\nu\beta\beta$ process as mentioned above. Hence, the same set of model parameters which are supposed to produce correct neutrino phenomenology can also be used to estimate the observables of different low energy processes. Thus this model is constrained from these processes also. The motivation is to see if the neutrino mass matrix that can explain the neutrino phenomenology can also provide sufficient parameter space for other low energy observables $0\nu\beta\beta$, cLFV etc. We also correlate the sterile neutrino mass with $0\nu\beta\beta$ and cLFV processes to see the impact of sterile neutrino. 

\begin{figure}[H]
	\begin{center}
		\includegraphics[width=0.45\textwidth]{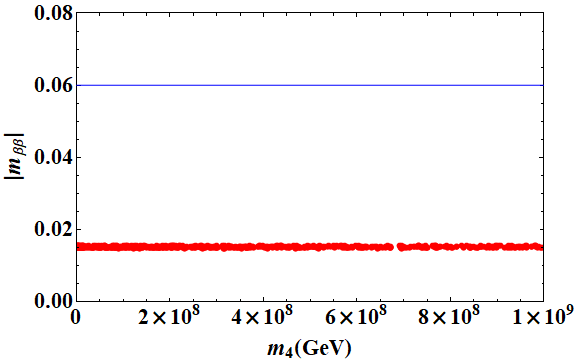}
		\includegraphics[width=0.45\textwidth]{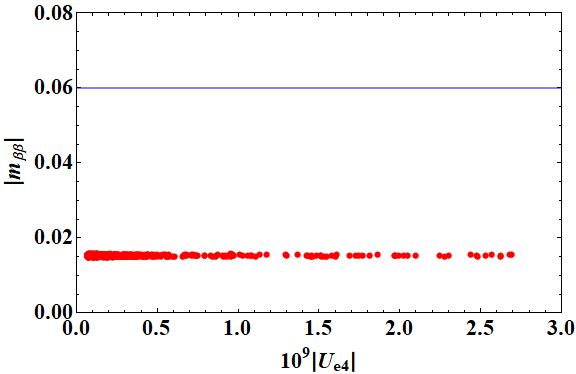}
	\end{center}
	\begin{center}
		\caption{The prediction of the effective neutrino mass as a function of sterile neutrino mass and mixing.}
		\label{fig2}
	\end{center}
\end{figure}

\begin{figure}[H]
	\begin{center}
		\includegraphics[width=0.45\textwidth]{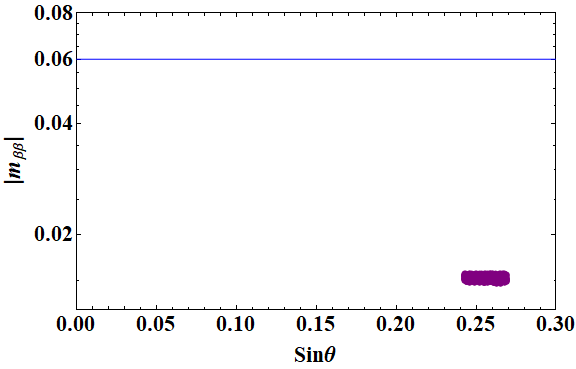}
	\end{center}
	\begin{center}
		\caption{The prediction of the effective neutrino mass as a function of $TM_{1}$ mixing parameter $Sin\theta$.}
		\label{fig2a}
	\end{center}
\end{figure}

\begin{figure}[H]
	\begin{center}
		\includegraphics[width=0.45\textwidth]{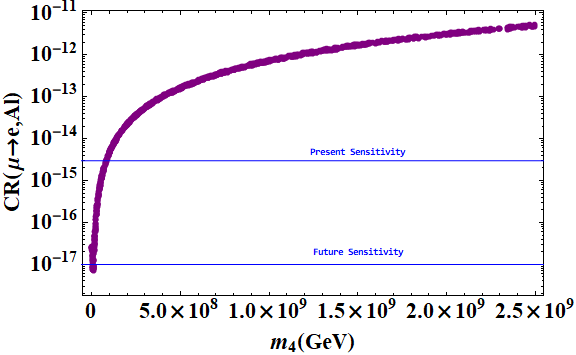}
		\includegraphics[width=0.45\textwidth]{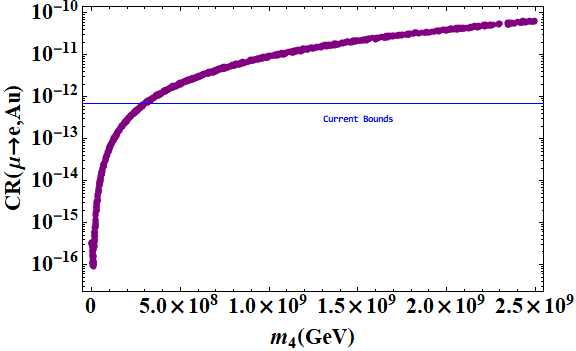}
	\end{center}
	\begin{center}
		\caption{$\text{CR}(\mu-e,N)$ as a function of sterile neutrino mass with two different targets. The blue horizontal line represents the experimental bounds on this process.}
		\label{fig5}
	\end{center}
\end{figure}

\begin{figure}[H]
	\begin{center}
		\includegraphics[width=0.45\textwidth]{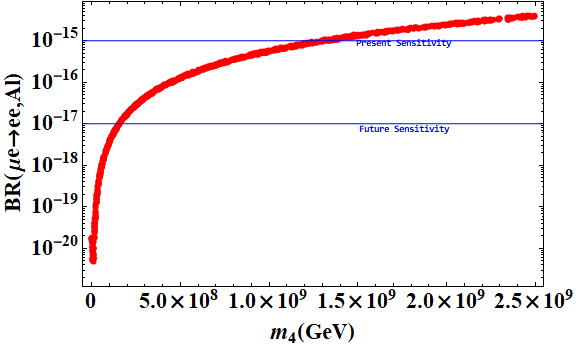}
		\includegraphics[width=0.45\textwidth]{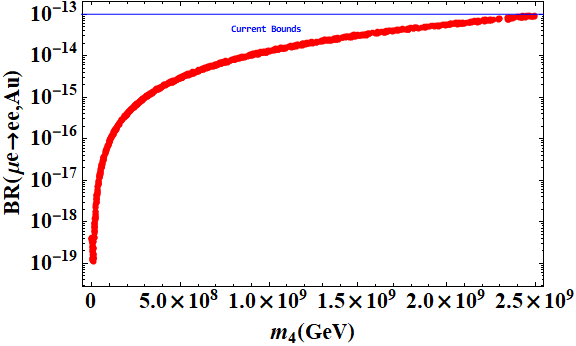}
	\end{center}
	\begin{center}
		\caption{$\text{BR}(\mu^{-}e^{-}\longrightarrow e^{-}e^{-},N)$ as a function of sterile neutrino mass with two different targets. The blue horizontal line represents the experimental bounds on this process.}
		\label{fig6}
	\end{center}
\end{figure}

The effective mass ($m_{\beta\beta}$) characterizing $0\nu\beta\beta$ process along with the presence of heavy sterile neutrino is calculated using Eq.(\ref{eq:21}). Fig \ref{fig2} shows the effective mass against the sterile neutrino mass and mixing. For new physics contribution coming from extra sterile neutrino, the effective mass is consistent with the upper bound ($|m_{\beta\beta}|\leq 0.06 eV$) followed from the data of KamLAND-ZEN \cite{KamLAND-Zen:2016pfg} experiment. It has been observed that the presence of sterile neutrino in the model results in a effective mass larger than that coming from the standard contribution (in the absence of sterile neutrino). However, even in the presence of such heavy sterile neutrino, the effective mass satisfies the experimental limit as the mixing of sterile neutrino with active neutrinos decreases with increase in mass in the model. Fig \ref{fig2a} shows variation of effective mass as a function of the parameter $\theta$ characterizing the $TM_{1}$ mixing. This plot shows how the model with $TM_{1}$ mixing constrains effective neutrino mass $m_{\beta\beta}$. 
We have performed the analysis of $\mu-e$ conversion with two different targets- Aluminium (Al) and Gold (Au). Fig \ref{fig5} shows the calculated conversion ratios with these two target as a function of the mass of the sterile neutrinos. In both the cases the results are within the reach of current and future experiments. It has been observed that sterile neutrino with mass $10^{8}$ GeV can lead to such process within the experimental bound.  
We have seen the sterile neutrino contribution to process $\mu^{-}e^{-}\longrightarrow e^{-}e^{-}$ in the model. Fig \ref{fig6} shows the variation of branching ratios with the mass of the sterile neutrinos. It has been observed that for targets with Al the experimental limits are reached for lower value of mass of the sterile neutrinos (around $10^{8}$ GeV) than in case with Au (around $2.5\times10^{9}$ Gev). This shows that the cLFV process induced by an additional sterile neutrinos could certainly be probed in near future experiments with Aluminium targets. The stringent bound on sterile neutrino mass to cause such process is around $3\times10^{8}$ GeV.

\begin{figure}[H]
	\begin{center}
		\includegraphics[width=0.55\textwidth]{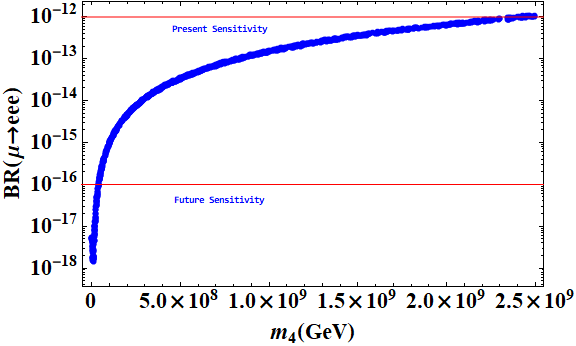}
		
	\end{center}
	\begin{center}
		\caption{$\text{BR} (\mu-eee)$ as a function of sterile neutrino mass. The red horizontal line represents the experimental bounds on this process.}
		\label{fig7}
	\end{center}
\end{figure}

\begin{figure}[H]
	\begin{center}
		\includegraphics[width=0.55\textwidth]{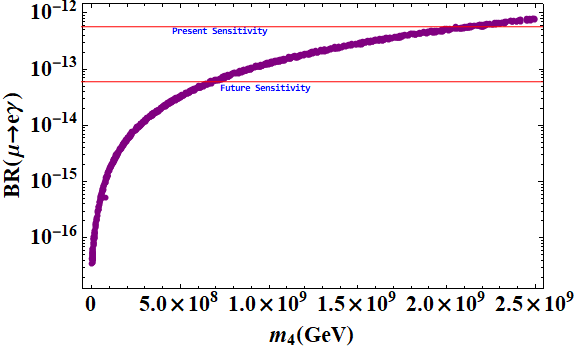}
		
	\end{center}
	\begin{center}
		\caption{$\text{BR} (\mu-e\gamma)$ as a function of sterile neutrino mass. The red horizontal line represents the experimental bounds on this process.}
		\label{fig7a}
	\end{center}
\end{figure}

Fig \ref{fig7} indicates the impacts of sterile neutrino in $\mu\longrightarrow eee$ process. It is evident from the figure that the branching ratios have a stronger experimental potential, with contributions well within current (future) experimental reach for sterile masses above $2\times10^{9}$ ($10^{8}$) GeV.
The branching ratios of another appealing process $\mu\longrightarrow e\gamma$ in presence of heavy sterile neutrino as a function of its mass is shown in fig \ref{fig7a}. It is seen that the results are well within current (future) experimental reach for sterile masses above $2\times10^{9}$ ($10^{9}$) GeV.

Similarly,we have carried out our analysis for processes involving tau atoms and calculated the observables using Eq.(\ref{eq:10a},\ref{eq:10b},\ref{eq:10c}).The results are shown in the following figs \ref{fig8},\ref{fig9},\ref{fig10}. It is observed that the sterile neutrino can have sizable contributions to such processes only when it has mass above $10^{9}$ GeV which is quite higher than that in case of processes involving muonic atoms. For the process $\tau\longrightarrow e\gamma$, the current experimental bound on branching ratio is achieved for $m_{s}>2\times10^{12}$ GeV, however lower mass of sterile neutrino (around $10^{12}$) can contribute to such process in future experiments as shown in fig \ref{fig8}. Fig \ref{fig9} indicates that the contributions of sterile neutrino in the process $\tau\longrightarrow \mu\gamma$  which are well within the current experimental limit for $m_{s}>2\times10^{12}$ GeV and the sensitivity of future experiments is reached for lower mass of sterile neutrino (around $5\times10^{11}$). For the process $\tau\longrightarrow eee$, the current and future experimental bound on branching ratio is achieved for $m_{s}>10^{12}$ GeV and around $3\times10^{11}$ respectively which can be seen in fig \ref{fig8}. In the table (\ref{tab5}), we have summarised the constraints on sterile neutrino mass coming from different cLFV processes.
\begin{figure}[H]
	\begin{center}
		\includegraphics[width=0.45\textwidth]{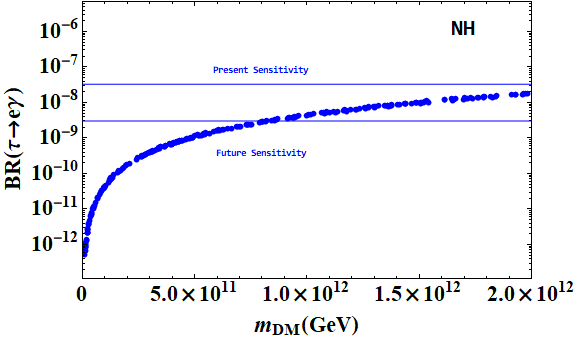}
		
	\end{center}
	\begin{center}
		\caption{$\text{BR} (\tau-e\gamma)$ as a function of sterile neutrino mass. The blue horizontal line represents the experimental bounds on this process.}
		\label{fig8}
	\end{center}
\end{figure}
\begin{figure}[H]
	\begin{center}
		\includegraphics[width=0.45\textwidth]{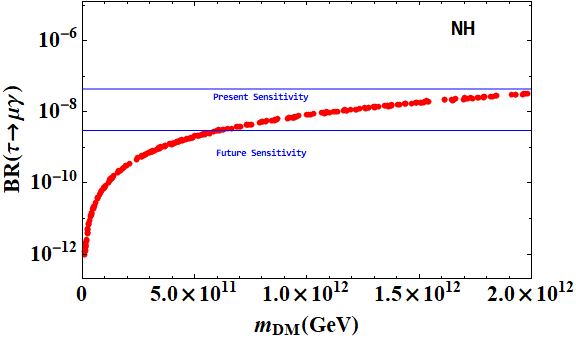}
	\end{center}
	\begin{center}
		\caption{$\text{BR}(\tau-\mu\gamma)$  as a function of sterile neutrino mass. The blue horizontal line represents the experimental bounds on this process.}
		\label{fig9}
	\end{center}
\end{figure}
\begin{figure}[H]
	\begin{center}
		\includegraphics[width=0.45\textwidth]{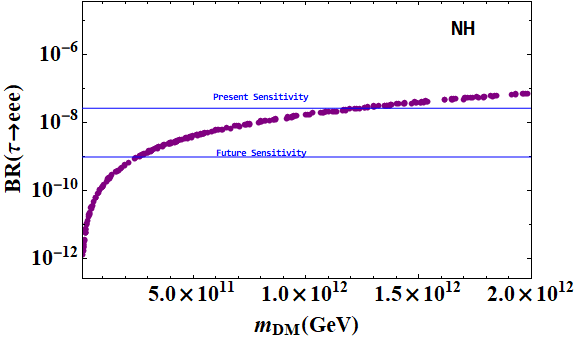}
		
	\end{center}
	\begin{center}
		\caption{$\text{BR} (\tau-eee)$  as a function of sterile neutrino mass. The blue horizontal line represents the experimental bounds on this process.}
		\label{fig10}
	\end{center}
\end{figure}

\begin{table}[H]
	\centering
	\begin{tabular}{|c|c|}
		
		\hline 
	cLFV Process	& Bounds on sterile neutrino mass  \\ 
		\hline 
	$(\mu e\longrightarrow ee,Al)$ & $3\times10^{8}$\\
		\hline
	$(\mu e\longrightarrow ee,Au)$ & $2.5\times10^{9}$\\
		\hline
	$\mu\longrightarrow eee$	& $10^{8}$  \\ 
		\hline
	$\mu\longrightarrow e\gamma$ & $10^{9}$  \\ 
		\hline
	$(\mu - e,Al)$ & $2\times10^{8}$\\ 
		\hline
	$(\mu - e,Au)$ & $5\times10^{8}$\\ 
		\hline
	$\tau\longrightarrow e\gamma$ &  $10^{12}$   \\ 
		\hline
	$\tau\longrightarrow \mu\gamma$& $5\times10^{11}$ \\ 
		\hline
    $\tau\longrightarrow eee$& $3\times10^{11}$ \\ 
		\hline  
	\end{tabular} 
	\caption{Constraints on sterile neutrino mass from different cLFV processes.}\label{tab5}
\end{table} 
\section{\label{sec:level6}Conclusion}
In this work, we have studied the effect of sterile neutrino on the low energy processes focusing on charged lepton flavor violation and neutrinoless double beta decay. The framework of our study is an MES model which is obtained by the addition of a triplet of right handed neutrinos and a sterile neutrino singlet field to the standard model. The gauge group of standard model is extended by the flavor symmetry group $\Delta(96)$ along with two $C_{2}$ groups and one $C_{3}$ group. The model is constructed in such a way that it gives rise to a special mixing pattern known as $TM_{1}$ mixing. The model leading to $TM_{1}$ mixing with $\mu \text{-}\tau$ symmetry predicts maximal atmospheric mixing angle and maximal breaking of the CP symmetry. These two important constraints of the model comply with the experimental data. Moreover, our construction of the model rules out inverted ordering of the neutrino masses. 
The model is represented by three model parameters that have been evaluated by comparing the light neutrino oscillation parameters in $3\sigma$ range. We have obtained the sterile neutrino mass and mixing from the model parameters. We then feed the model parameters in calculating different observables characterizing different low energy processes. Sizeable implications for 0$\nu\beta\beta$ can be obtained within the MES model with $TM_{1}$ mixing. The texture of the mass matrices predict the effective mass $m_{\beta\beta}$ that is consistent with the experimental data. We have investigated different cLFV processes involving muon and tau leptons. It has been observed that wide range of parameter space has the possibility to be probed in near future experiments. There are no theoretical upper bounds on the mass of the sterile neutrino. However, in this model the different cLFV processes highly constrain the mass of sterile neutrino. In this work, we have summarized the limits on the mass of the sterile neutrino to contribute such processes. Another important conclusion that can be drawn from the present work is that the sterile neutrino mass range allowed by different cLFV processes can give rise to effective neutrino mass within the experimental limits. Thus the two low energy observables can also be correlated in the proposed model.

In conclusion, the MES model with $\Delta96$ discrete flavor symmetry can address neutrino phenomenology in presence of heavy sterile neutrino with the prediction of experimentally observed neutrino parameters. We have shown that the model have interesting implications in rare decay experiments like lepton flavor violation and also neutrinoless double beta decay. The estimation of the model on baryon asymmetry of the universe (BAU) can also be studied in future.
\appendix
\section{Properties of $\Delta(96)$ group} 
\label{appen1}

$\Delta96$ is one of the members of $\Delta6n^{2}$ with $n=4$.
P,Q and C are the generators of $\Delta96$ which can be given as,
\begin{equation} \label{eq:u} 
P = \left(\begin{array}{ccc}
0 & 0 & 1 \\
0 & -1 & 0 \\ 
1 & 0 & 0 \\ 
\end{array}\right),\;  Q  = \left(\begin{array}{ccc}
0 & 1 & 0 \\
0 & 0 & 1\\ 
1 & 0 & 0
\end{array}\right), \; C =\left(\begin{array}{ccc}
1 & 0 & 0 \\
0 & i & 0\\ 
0 & 0 & -i
\end{array}\right).
\end{equation}
There are $11$ irreducible representations of $\Delta(96)$ , two singlets 1 and $1^{\prime}$, one doublet 2, 
six triplets 3, $3^{\prime}$,$3_{i}$,$3_{i}^{\prime}$,$\bar{3_{i}}$, $\bar{3_{i}^{\prime}}$ and $6$. We note that the first five representations correspond to that of $S_{4}$ which is a subgroup of $\Delta96$. The character table for $\Delta96$ is given below in \ref{tabA1}  
\begin{table}[H]
	\centering
	\begin{tabular}{|c|c|c|c|c|c|c|c|c|c|c|}
	    \hline 
	    $\Delta(96)$ &$C_{0}$  & $C_{1}$ & $C_{2}$ & $C_{3}$ & $C_{4}$& $C_{5}$ &$C_{6}$ &$C_{7}$ &$C_{8}$ &$C_{9}$ \\
		\hline 
		1 & 1 &1 &1& 1 & 1 & 1 & 1 &1&1&1\\
		\hline 
		$1^{\prime}$& $1$ & $1$ & $-1$ &$-1$& $1$ & $-1$ & $-1$ &  $1$ &$1$ &$1$ \\
		\hline 
   	    2& 2& 2 &0& 0 & $-1$ &0 &  0 &2 &2 & 2  \\
	    \hline 
		3& 3 & 3 &1& 1 & 0 &-1 &  -1 &-1 & -1 & -1  \\
		\hline 
        $3^{\prime}$& 3 & 3 &-1& -1 & 0 &1 &  1 &-1 & -1 & -1   \\
		\hline 
		$3_{i}$& 3 & -1 & i& -i & 0 &-1 &  1 &1 & $z$ & $\bar{z}$   \\
		\hline
		$\bar{3_{i}}$& 3 & -1 & -i& i & 0 &-1 &  1 &1 &$\bar{z}$ & $z$   \\
		\hline
		$3_{i}^{\prime}$& 3 & -1 & -i& i & 0 &1 &  -1 &1 & $z$ & $\bar{z}$   \\
		\hline
		$\bar{3_{i}^{\prime}}$& 3 & -1 & i& -i & 0 & 1 &  -1 &1 & $\bar{z}$ & $z$    \\
		\hline
		$6$& 0 & -2 & 0& 0 & 0 &0 &  0 &-2 & 2 & 2   \\
		\hline
	\end{tabular} 
	\caption{Character table of $\Delta96$ group.}\label{tabA1}
\end{table}
Here, The tensor products of 1, $1^{\prime}$, 2, 3, $3^{\prime}$ follow the product rules of $S_{4}$ 
\begin{equation}
3 \times 1 = 3,  3 \times 1^{\prime} = 3^{\prime},  3^{\prime} \times 1^{\prime} = 3 ,  2 \times 1^{\prime}=2.\\
\end{equation}
\begin{gather}
2 \otimes 3 = 3 \oplus  3^{\prime},\\
3 \otimes 3 = 1_{1} \oplus 2 \oplus 3 \oplus 3^{\prime},\\
3^{\prime} \otimes 3^{\prime}= 1_{1} \oplus 2 \oplus 3 \oplus 3^{\prime}.\\
3_{i} \otimes 3_{i}= 3^{\prime} \oplus \bar{3_{i}^{\prime}} \oplus \bar{3_{i}}\\
\bar{3_{i}} \otimes 3_{i}= 1 \oplus 2 \oplus 6
\end{gather}
For the Clebsch-Gordon coefficients all the above expansion, please refer \cite{King:2013vna,King:2014rwa}. The tensor products involving $3_{i}$ and $\bar{3_{i}^{\prime}}$ are given by,    
\begin{gather}
\left(\begin{array}{c}
a_{1}\\
a_{2}\\
a_{3} \end{array}\right)_{3_{i}}\otimes \left(\begin{array}{c}
b_{1}\\
b_{2}\\
b_{3} \end{array}\right)_{3_{i}} = \left(\begin{array}{c}
a_{1}b_{1}\\
a_{2}b_{2}\\
a_{3}b_{3}
\end{array}\right)_{3^{\prime}} \oplus \left(\begin{array}{c}
a_{2}b_{3}+a_{3}b_{2}\\
a_{2}b_{3}+a_{3}b_{2}\\
a_{1}b_{2}+a_{2}b_{1}
\end{array}\right)_{\bar{3_{i}^{\prime}}} \oplus  \\
\left(\begin{array}{c}
a_{2}b_{3}-a_{3}b_{2}\\
a_{2}b_{3}-a_{3}b_{2}\\
a_{1}b_{2}-a_{2}b_{1} \end{array}\right)_{\bar{3_{i}}}.
\end{gather}
\begin{gather}
\left(\begin{array}{c}
a_{1}\\
a_{2}\\
a_{3} \end{array}\right)_{3_{i}}\otimes \left(\begin{array}{c}
b_{1}\\
b_{2}\\
b_{3} \end{array}\right)_{\bar{3_{i}}} = (a_{1}b_{1}+a_{2}b_{2}+a_{3}b_{3})_{1} \oplus \left(\begin{array}{c}
1/\sqrt{2}(a_{2}b_{2} - a_{3}b_{3})\\
1/\sqrt{6}(-2a_{1}b_{1}+ a_{2}b_{2} + a_{3}b_{3})
\end{array}\right)_{2} \oplus  \\
\left(\begin{array}{c}
a_{2}b_{3}\\
a_{3}b_{1}\\
a_{1}b_{2} \\
a_{3}b_{2}\\
a_{1}b_{3}\\
a_{2}b_{1} \end{array}\right)_{6}.
\end{gather}
\section*{Acknowledgements}
NG would like to acknowledge Department of Science and Technology (DST),India(grant DST/INSPIRE Fellowship/2016/IF160994) for the financial assistantship. MKD acknowledges the Department of Science and Technology, Government of India for the support under the project no. $EMR/2017/001436$.

\bibliographystyle{paper}
\bibliography{LFV}
\end{document}